\documentclass[prb,showpacs,twocolumn]{revtex4}

\usepackage{epsfig}

\usepackage{graphics}

\usepackage{graphicx}

\usepackage{dcolumn}

\usepackage{bm}

\newcommand{\sba}{\begin{subeqnarray}}

\newcommand{\sea}{\end{subeqnarray}}

\def\cm-1{cm$^{-1}$}

\begin{document}

\title{Diagrammatic analysis of the Hubbard model:\\Stationary property of the thermodynamic potential}
\author{V.\ A.\ Moskalenko$^{1,2}$}
\email{mosckalen@theor.jinr.ru}
\author{L.\ A.\ Dohotaru$^{3}$}
\author{I.\ D.\ Cebotari$^{1}$}
\affiliation{$^{1}$Institute of Applied Physics, Moldova Academy
of Sciences, Chisinau 2028, Moldova} \affiliation{$^{2}$BLTP,
Joint Institute for Nuclear Research, 141980 Dubna, Russia}
\affiliation{$^{3}$Technical University, Chisinau 2004, Moldova}
\date{\today}
\begin{abstract}
%
Diagrammatic approach proposed many years ago for strong correlated Hubbard model is
developed for analyzing of the thermodynamic
potential properties.
The new exact relation between such renormalized quantities as
thermodynamic potential, one-particle propagator and correlation
function is established. This relation contains additional
integration of the one-particle propagator by the auxiliary
constant. The vacuum skeleton diagrams constructed from
irreducible Green's functions and tunneling propagator lines are
determined and special functional is introduced. The properties of
such functional are investigated and its relation to the
thermodynamic potential is established. The stationary properties
of this functional with respect to first order changing of the
correlation function is demonstrated and as a consequence the
stationary properties of the thermodynamic potential is proved.
\end{abstract}

 \maketitle

\section{Introduction}

The Hubbard model is one of the most important models for the
electron of solids which describes quantum mechanical hopping
of electron between lattice sites and their short ranged
repulsive Coulomb interaction.

This model was discussed by Hubbard $^{[1]}$ in order to
describe a narrow-band system of transition metals and has been
revised to investigate the properties of highly correlated
electron systems such as the copper oxide superconductors and
others.

Hubbard model exhibits various phenomena including
metal-insulator transition, antiferromagnetism, ferromagnetism
and superconductivity. This model assumes that each atom of the
crystal lattice has only one electron orbit and  the
corresponding orbital state is non degenerate.

The Hamiltonian of Hubbard model is a sum of the two terms
%
\begin{eqnarray}
H&=&H^{0}+H^{\prime},  \label{1}
\end{eqnarray}
where $H^{0}$ is the atomic contribution, which contains the Coulomb
interaction term $U$ and local electron energy $\overline{\epsilon}$ on the
atom
%
\begin{eqnarray}
H^{0}&=&\sum\limits_{\overrightarrow{i}}H_{\overrightarrow{i}}^{0},  \nonumber \\
H_{\overrightarrow{i}}^{0}&=&\sum\limits_{\sigma }\epsilon n_{\overrightarrow{i}\sigma }
+Un_{\overrightarrow{i}\uparrow }n_{\overrightarrow{i}\downarrow }, \\
\epsilon&=&\overline{\epsilon}-\mu,\quad
n_{\overrightarrow{i}\sigma}=C_{\overrightarrow{i}\sigma }^{+}C_{\overrightarrow{i}\sigma }  \nonumber \label{2}
%
\end{eqnarray}
and hopping Hamiltonian
%
\begin{eqnarray}
H^{\prime}&=&\sum\limits_{\overrightarrow{i}\overrightarrow{j}}\sum\limits_{\sigma}t(\overrightarrow{i}-\overrightarrow{j})C_{\overrightarrow{i}\sigma
}^{+}C_{\overrightarrow{j}\sigma },\\  \nonumber
t(\overrightarrow{i}-\overrightarrow{j})&=&t^{*}(\overrightarrow{j}-\overrightarrow{i}),\quad
t(0)=0.\label{3}
%
\end{eqnarray}
Here $C_{\overrightarrow{i}\sigma }^{+}(C_{\overrightarrow{i}\sigma })$ are the creation
(annihilation) electron operators with local site $\overrightarrow{i}$ and spin
$\sigma$.
Because in the thermodynamic perturbation theory we
shall use thermal averages in a grand canonical ensemble we have
added to the Hamiltonian (1) the term $-\mu\hat{N}_{e}$
%
\begin{eqnarray}
\hat{N}_{e}&=&\sum\limits_{\overrightarrow{i}\sigma}n_{\overrightarrow{i}\sigma }\label{4}
%
\end{eqnarray}
where $\mu$ is the chemical potential and $\hat{N}_{e}$
electron number operator. The quantities $U$ and $\hat{N}_{e}$
are the fundamental parameters of the model and because of
large value of the Coulomb repulsion it is taken into account
in zero approximation of our theory. The operator $H^{\prime}$,
which describes hopping of the electrons between sites of the
crystal lattice is regarded as a perturbation.

For investigating this model new physical and mathematical
concepts and techniques have been elaborated. We can only
enumerate some of them. There are many analytical
approximations as Hubbard approximation, noncrossing
approximation (NCA), slave-boson method, Dynamical Mean Field
Theory (DMFT), composite and others projection operator methods
and approaches. Also numerical simulations of thermodynamic
quantities and density of states have been performed by various
methods. Some exact results for one- and two-dimensional
Hubbard model are known. Each approximate method has its
advantages and disadvantages. A short and comprehensive reviews
of the methods can be found in papers and books $^{[2-7]}$.
Besides these approaches we must enumerate also the special
diagram techniques elaborated for strongly correlated electron
systems. Stasyuk $^{[8]}$, Zaitsev $^{[9]}$, Izyumov $^{[10]}$
and coworkers have developed a diagram technique for Hubbard
model, based on the disentangling of nondiagonal Hubbard
operator $\chi{mn}$ out of a time-ordered products of other
such operators. Due to a more complicate algebra of Hubbard
transfer operator than that for Fermi operators the essential
features of this technique remain till now poorly developed.

The other diagrammatic approach around the atomic limit also
has been proposed for Hubbard model both in normal $^{[11,12]}$
and in superconducting  state $^{[13]}$. This theory introduces
the Generalized Wick Theorem (GWT) that uses cumulant expansion
of the statistical average values for the products of the
Fermion operators. The GWT takes into account the fact that
Hamiltonian $H^{0}$ is nonquadratic in fermion operators due to
Coulomb interaction. This last circumstance is responsible for
the appearance of the nonvanishing site cumulants called
irreducible Green's functions. These new  Green's functions
take into account all the spin, charge and pairing fluctuations
of the system. The perturbation formalism around atomic limit
has the advantage that local (atomic) physical properties can
be evaluated exactly and that the transfer of this local
information to neighboring sites due to the kinetic mobility of
the conduction electrons can be handled perturbatively  in
powers of hopping integral.

A GWT for chronological averages of products of electron operators
was formulated later also by Metzner $^{[14]}$. Metzner did not
derive a Dyson-type equation for the renormalized one-particle
Green's function and the role of our correlation function has not
established.

In spite of the similarity to our works, the diagrams in his
approach are quite different. The $n$ - particle cumulant is
represented a $2n$ - valent-point vertex with attached $n$ entering
and $n$ leaving lines, where as our such a cumulant or irreducible
Green function is represented by a rectangle with $2n$ vertices with
$n$ entering  and $n$ leaving arrows. The rectangle contours the
vertices with the same site index but different time and spin
labels. In addition Metzner investigated the limit of high lattice
dimensions.

In this paper we shall develop the diagrammatic theory proposed
before for Hubbard model $^{[11,12]}$ with the aim to demonstrate
the existence of the relation between renormalized quantities of
thermodynamic potential and one-particle Green's function and also
to prove the stationary properties of this potential.

Such theorem was proved firstly by Luttinger and Ward $^{[15]}$ for
uncorrelated systems  by using the diagrammatic technique of weak
coupling field theory.

The strong coupling diagram theory used by us needs new conceptions
and new equations and they are used to prove stationary property of
thermodynamic potential for strongly correlated systems. Such a
proof has been already achieved for Anderson impurity model in paper
$^{[16]}$.

The paper is organized in the following way. In section II we
develop the diagrammatic theory in the strong coupling limit
and the skeleton diagrams are introduced. In Section III we
prove the stationary theorem for renormalized thermodynamic
potential. Section IV has the conclusions.


\section{Perturbation Treatment}

%
We shall use the definition of the one-particle Matsubara
Green's functions in interaction representation as in paper
$^{[11,12]}$:
\begin{eqnarray}
%
G(x|x^{\prime})&=&-\ \left\langle
TC_{\overrightarrow{x}\sigma}(\tau)\overline{C}_{\overrightarrow{x}^{\prime}\sigma^{\prime}}
(\tau^{\prime})U(\beta)\right\rangle _{0}^{c}, \label{5}
%
\end{eqnarray}
%
where  $x$ stands for $(\overrightarrow{x},\sigma,\tau)$ and
index $c$ for $\left\langle ...\right\rangle _{0}^{c}$ means
the connected part of the diagrams which appear in the
right-hand of part of (5). We use the series expansion for the
evolution operator $U(\beta)$ with some generalization because
we introduce the auxiliary constant $\lambda$ and use $\lambda
H^{\prime}$ instead of $H^{\prime}$:
%
\begin{eqnarray}
%
U_{\lambda}(\beta ) & = & T\exp (-\lambda\int\limits_{0}^{\beta }
H^{\prime}(\tau )d\tau ).\label{6}
%
\end{eqnarray}
%
In the presence of this constant we shall use index $\lambda$
for all dynamical quantities as $G_{\lambda}(x|x^{\prime})$ and
so on. At the last stage of the calculations this constant will
be put equal to one.

In zero order approximation the one-particle Green's function
$G^{(0)}(x|x^{\prime})$ is local
\begin{eqnarray}
%
G^{(0)}(x|x^{\prime})&=&\delta_{\overrightarrow{x}\overrightarrow{x}^{\prime}}
\delta_{\sigma\sigma{\prime}}G^{(0)}(\tau-\tau^{\prime}) \label{7}
%
\end{eqnarray}
%
and the Fourier representation of the last function is
%
\begin{eqnarray}
%
G_{\sigma }^{(0)}(i\omega_{n} )&=&\int\limits_{0}^{\beta
}G_{\sigma }^{(0)}(\tau)e^{i\omega_{n}\tau}d\tau \nonumber\\
&=&\frac{1}{Z_{0}}\{\frac{e^{-\beta E_{0}}+e^{-\beta
E_{\sigma}}}{i\omega_{n} +E_{0}-E_{\sigma}}+\frac{e^{-\beta
E_{\overline{\sigma }}}+e^{-\beta E_{2}}}{i\omega_{n}
+E_{\overline{\sigma}}-E_{2}}\},\nonumber\\
&&\\
Z_{0}&=&e^{-\beta E_{0}}+e^{-\beta E_{\sigma}}+e^{-\beta
E_{\overline{\sigma }}}+e^{-\beta E_{2}},\nonumber\\
E_{0}&=&0,\quad E_{\sigma}=E_{\overline{\sigma
}}=\epsilon,\quad E_{2}=2\epsilon+U,\nonumber\\ \omega_{n}
&=&(2n+1)\pi /\beta.\nonumber\label{8}
%
\end{eqnarray}
%
Here $E_{0},E_{\sigma}$ and $E_{2}$ are the energies of atomic
sites.

As has been proved in papers $^{[11,12]}$ propagator (5) has
the diagrammatic representation depicted on the Fig. 1.
%
\begin{figure*}[t]
%
\centering
\includegraphics[width=0.85\textwidth,clip]{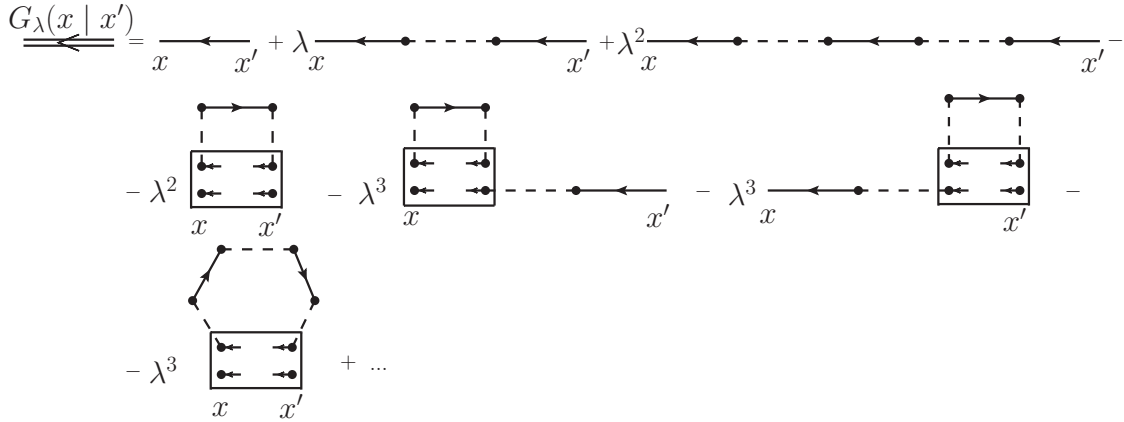}
\vspace{-0mm}
%
\caption{First three orders of perturbation theory for
one-particle propagator. The thin solid line depicts zero order
propagator, thin dashed line depicts the tunneling matrix element.
The rectangles depict irreducible two-particles Green's functions
$G_{2}^{(0)ir}$ and points are the vertices of
diagrams.}\label{fig-1} \vspace{-5mm}
\end{figure*}
%
The irreducible Green's functions of order $n$ are depicted with
rectangles with $2n$ vertices. The arrows which enter in the
vertex point depict annihilation electrons and that which go out
the created electrons.

In paper $^{[11]}$ we have introduced the notion of correlation
function $Z_{\lambda}(x|x^{\prime})$ which is the sum of
strongly connected diagrams containing irreducible Green's
functions and related to a more convenient function
$\Lambda_{\lambda}(x|x^{\prime})=G^{0}(x|x^{\prime})+Z_{\lambda}(x|x^{\prime})$.

In Fig. 1 the fourth and seventh diagrams of the right-hand
site belong to correlation function. Due to the fact that
irreducible functions are local and tunneling matrix elements
have the property  $t(\overrightarrow{x}-\overrightarrow{x})=0$
in Fig. 1 are omitted all the diagrams which contain
self-locked tunneling elements.

As is seen from Fig. 1 the process of propagator
renormalization is accompanied by the analogous process for
tunneling matrix elements renormalization and replacing of the
instant quantity $\lambda t(x-x^{\prime})=\lambda
t(\overrightarrow{x}-\overrightarrow{x}^{\prime})\delta(\tau-\tau^{\prime}-0^{+}))$
by dynamical one $\widetilde{T}_{\lambda}(x|x^{\prime})$ equal
to
\begin{eqnarray}
\widetilde{T}_{\lambda}(x|x^{\prime})&=&\lambda
t(\overrightarrow{x}-\overrightarrow{x}^{\prime}) \delta(\tau-\tau^{\prime}-0^{+})\nonumber\\
&+&\sum\limits_{12}\lambda t(x-1)G_{\lambda}(1|2)\lambda t(2-x^{\prime})\label{9}
\end{eqnarray}
which in Fourier representation
%
\begin{eqnarray}
t(\overrightarrow{x})&=&\frac{1}{N}\sum\limits_{\overrightarrow{k}}
\epsilon(\overrightarrow{k})\exp{(-i\overrightarrow{k}\overrightarrow{x})},\nonumber\\
G_{\lambda}(x|x^{\prime})&=&\frac{1}{N}\sum\limits_{\overrightarrow{k}}\frac{1}{\beta}
\sum\limits_{\omega_{n}}G_{\lambda}(\overrightarrow{k}|i\omega_{n})\nonumber\\
&\times&
\exp{[-i\overrightarrow{k}(\overrightarrow{x}-\overrightarrow{x}^{\prime})
-i\omega_{n}(\tau-\tau^{\prime})]}\nonumber
%
\end{eqnarray}
has a form:
\begin{eqnarray}
\widetilde{T}_{\lambda}(\overrightarrow{k}|i\omega_{n})&\equiv&\lambda
T_{\lambda}(\overrightarrow{k}|i\omega_{n})\nonumber\\
&=&\lambda \epsilon(\overrightarrow{k})(1+\lambda \epsilon(\overrightarrow{k})
G_{\lambda}(\overrightarrow{k}|i\omega_{n}))\label{10}
\end{eqnarray}

The renormalized tunneling matrix element $T_{\lambda}$ really is tunneling
Green's function and will be depicted as double dashed line.
$\widetilde{T}_{\lambda}$ is represented by such double dashed line multiplied by $\lambda$.

Now we introduce the skeleton diagrams which contain only
irreducible Green's functions and simple dashed lines without
any renormalization. In such skeleton diagrams the thin dashed
lines are replaced by double dashed lines with realizing the
complete renormalization of dynamical quantities.

The skeleton diagrams for correlation $\Lambda_{\lambda}$
function are depicted on Fig. 2.

As can bee seen from Fig. 2 the skeleton diagrams for $\Lambda(k)$
function are of two kinds. The first four diagrams of this Figure
are local and therefore their Fourier representation don't depend on
momentum. The last diagram and other ones with more number of
rectangles are non localized and their Fourier representation
depends on momentum. Only the first category of diagram are taken
into account in Dynamical Mean Field Theory.
%
\begin{figure*}[t]
%
\centering
\includegraphics[width=0.85\textwidth,clip]{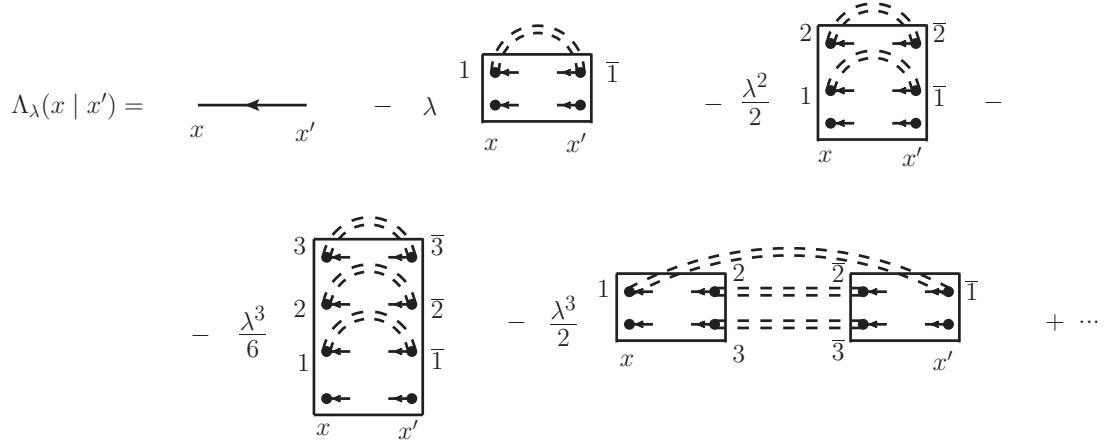}
\caption{ The skeleton diagrams for correlation function $\Lambda_{\lambda}$. Double
dashed lines depict the tunneling Green's functions $T_{\lambda}$ and the rectangles
depict the irreducible Green's functions.
}\label{fig-2} \vspace{-5mm}
%
\end{figure*}

As was proved in papers $^{[11,12]}$ the knowledge of this
function permits to formulate the following Dyson-type equation
for one-particle Green's function
%
\begin{equation}
%
G_{\lambda}(k)=\frac{\Lambda
_{\lambda}(k)}{1-\lambda\epsilon(\overrightarrow{k})\Lambda
_{\lambda}(k)}.\label{11}
%
\end{equation}
%
Here $k$ stands for $(\overrightarrow{k},i\omega_{n})$ with odd
Matsubara frequencies. Equation (10) and (11) gives us the results
\begin{eqnarray}
\widetilde{T}_{\lambda}(k)&=&\lambda
T_{\lambda}(k),\nonumber\\
T_{\lambda}(k)&=&\frac{\epsilon(\overrightarrow{k})}{1-\lambda\epsilon(\overrightarrow{k})\Lambda
_{\lambda}(k)}.\label{12}
\end{eqnarray}
Equation (12) has a form of Dyson equation for tunnelind Green's
function, and the role of mass operator $\Sigma_{\lambda}$ is
carried out by correlation  function multiplied by auxiliary
constant $\lambda$:
\begin{eqnarray}
\Sigma_{\lambda}(k)&=&\lambda \Lambda_{\lambda}(k). \label{13}
\end{eqnarray}

In Hubbard I approximation we neglect the correlation function
$Z_{\lambda}(k)$ and consider function $\Lambda_{\lambda}(k)$ equal
to zero order Green's function $G^{0}(k)$. In this approximation we
have
%
\begin{eqnarray}
%
G_{\lambda}^{I}(k)=\frac{G ^{0}(k)}{1-\lambda\epsilon(\bf{k}) G
^{0}(k)},\nonumber
%
\end{eqnarray}
%
which describes two Hubbard energy subbands distanced between them
for $U \neq 0$ and without the possibility to describe the
Mott-Hubbard transition.

\section{Thermodynamic potential diagrams}


The thermodynamic potential of the system is determined by the
connected part of the mean value of the evolution operator
$^{[11,12]}$
%
\begin{eqnarray}
%
F&=&F_{0}-\frac{1}{\beta }\left\langle U(\beta )\right\rangle
_{0}^{c}. \label{14}
%
\end{eqnarray}
%
Let us consider a more general quantity first
%
\begin{eqnarray}
%
F(\lambda)&=&F_{0}-\frac{1}{\beta }\left\langle
U_{\lambda}(\beta )\right\rangle _{0}^{c}, \label{15}
%
\end{eqnarray}
%
and put then $\lambda=1$.

By using the perturbation theory we have obtained the first orders
of diagrams for $\left\langle U_{\lambda}(\beta )\right\rangle
_{0}^{c}$, depicted in Fig. 3.

%
\begin{figure*}[t]
%
\centering
\includegraphics[width=0.85\textwidth,clip]{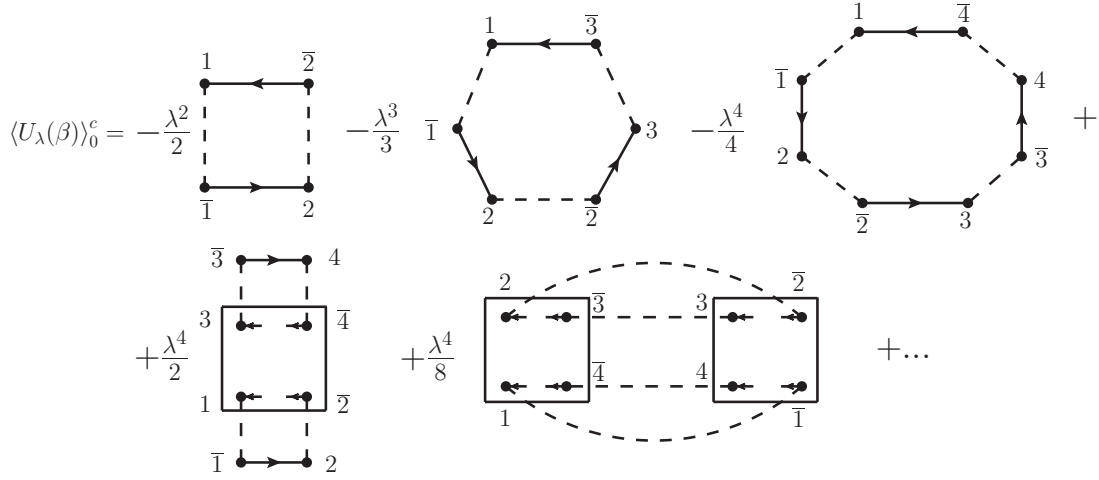}
\vspace{-0mm}
%
\caption{ The first four orders of perturbation theory for
$\left\langle U_{\lambda}(\beta )\right\rangle _{0}^{c}$.}
\label{fig-3} \vspace{-5mm}
\end{figure*}

In order to obtain the better understanding of these diagrammatic
contributions we examine the expression
%
\begin{eqnarray}
\sum\limits_{xx^{\prime}} G_{\lambda}(x|x^{\prime})\lambda
t(\overrightarrow{x}^{\prime}-\overrightarrow{x})
\delta(\tau-\tau^{\prime}-0^{+})\delta_{\sigma\sigma^{\prime}},\label{16}
%
\end{eqnarray}
where double repeated indices suppose summation and
integration. Consequently (16) is equal to
%
\begin{eqnarray}
&-&\beta\sum\limits_{\overrightarrow{x}\overrightarrow{x}^{\prime}}\sum\limits_{\sigma}
G_{\lambda\sigma}(\overrightarrow{x}-\overrightarrow{x}^{\prime}|-0^{+})\lambda
t(\overrightarrow{x}^{\prime}-\overrightarrow{x})\nonumber\\
&=&-\lambda\sum\limits_{\overrightarrow{k}\sigma}\sum\limits_{\omega_{n}}\epsilon(\overrightarrow{k})
G_{\lambda\sigma}(\overrightarrow{k}|i\omega_{n})\exp(i\omega_{n}0^{+}).
\label{17}
%
\end{eqnarray}
Here we have carried out the integration by time.

From diagrammatic point of view equation (16) implies the
procedure of locking of the external lines of propagators
$G_{\lambda}$ diagrams depicted on the Fig. 1 with the
tunneling matrix element
$t(\overrightarrow{x}^{\prime}-\overrightarrow{x})$ and
obtaining in such a way the diagrams without external lines
similar with ones for $\left\langle U_{\lambda}(\beta
)\right\rangle _{0}^{c}$ depicted on Fig. 3. These two series
of diagrams differ by coefficients in front of them.

In expression (17) the coefficients $\frac{1}{n}$ before each
diagram are absent, where $n$ is the order of perturbation
theory. These coefficients are present in Fig. 3. In order to
restore these $\frac{1}{n}$ coefficients in (17) and obtain the
coincidence with $\left\langle U_{\lambda}(\beta )\right\rangle
_{0}^{c}$ series it is enough to integrate by $\lambda$ the
expression (17) and obtain:
%
\begin{eqnarray}
-\sum\limits_{\overrightarrow{x}\overrightarrow{x}^{\prime}}\sum\limits_{\sigma}
\beta \int d\lambda
t(\overrightarrow{x}^{\prime}-\overrightarrow{x})
G_{\lambda\sigma}(\overrightarrow{x}-\overrightarrow{x}^{\prime}|-0^{+}).
\label{18}
%
\end{eqnarray}
The expression (18) displayed in a diagrammatic representation
coincides exactly with mean value of the evolution operator:
%
\begin{eqnarray}
\left\langle U_{\lambda}(\beta )\right\rangle
_{0}^{c}&=&-\sum\limits_{\overrightarrow{x}\overrightarrow{x}^{\prime}}
\beta
t(\overrightarrow{x}^{\prime}-\overrightarrow{x})\nonumber\\
&\times&\int\limits_{0}^{\lambda} d\lambda^{\prime}
G_{\lambda^{\prime}\sigma}(\overrightarrow{x}-\overrightarrow{x}^{\prime}|-0^{+}).
\label{19}
%
\end{eqnarray}
In Fourier representation we have
%
\begin{eqnarray}
\left\langle U_{\lambda}(\beta )\right\rangle
_{0}^{c}=-\int\limits_{0}^{\lambda}
d\lambda^{\prime}\sum\limits_{\overrightarrow{k}\sigma
\omega_{n}}\epsilon(\overrightarrow{k})
G_{\lambda^{\prime}\sigma}(\overrightarrow{k}|i\omega_{n})\exp(i\omega_{n}0^{+}).\nonumber
&&\\ \label{20}
%
\end{eqnarray}
From (15) and (20) we obtain
%
\begin{eqnarray}
F(\lambda)&=&F_{0}+\int\limits_{0}^{\lambda}
d\lambda^{\prime}\sum\limits_{\overrightarrow{k}\sigma}\frac{1}{\beta}\sum\limits_{\omega_{n}}\epsilon(\overrightarrow{k})\nonumber\\
&\times&G_{\lambda^{\prime}\sigma}(\overrightarrow{k}|i\omega_{n})\exp(i\omega_{n}0^{+}).
\label{21}
%
\end{eqnarray}
Using the definition (14), the equation (21) can be written in
the form:
%
\begin{eqnarray}
F(\lambda)&=&F_{0}+\int\limits_{0}^{\lambda}\frac
{d\lambda^{\prime}}{\lambda^{\prime}}\sum\limits_{\overrightarrow{k}\sigma
}\frac{1}{\beta}\sum\limits_{ \omega_{n}}
T_{\lambda^{\prime}}(k)\nonumber\\
&\times&\Sigma_{\lambda^{\prime}}(k)\exp(i\omega_{n}0^{+}).
\label{22}
%
\end{eqnarray}
From (22) we have
%
\begin{eqnarray}
\lambda\frac{F(\lambda)}{d\lambda}&=&\sum\limits_{\overrightarrow{k}\sigma
}\frac{1}{\beta}\sum\limits_{ \omega_{n}}
T_{\lambda}(k)\Sigma_{\lambda}(k)\exp(i\omega_{n}0^{+})\nonumber\\
&=&\frac{1}{\beta}Tr(T_{\lambda}\Sigma_{\lambda}). \label{23}
%
\end{eqnarray}
In order to have a full  system of equations we add to (22) the
definition of the chemical potential of the system
%
\begin{eqnarray}
N_{e}&=\sum\limits_{\overrightarrow{k}\sigma
}\frac{1}{\beta}\sum\limits_{ \omega_{n}}
G_{\sigma}(\overrightarrow{k}|i\omega_{n})\exp(i\omega_{n}0^{+}),
 \label{24}
%
\end{eqnarray}
where $N_{e}$ is the electron number.

Equation (21) establishes the relation between thermodynamic
potential and renormalized one-particle propagator. This last
quantity depends on the auxiliary parameter $\lambda$ and (21)
contains an additional integration over it and is awkward
because that.

We shall obtain more convenient equation for thermodynamic
potential without such integration by $\lambda$. For that we shall
introduce a special functional
%
\begin{eqnarray}
Y(\lambda)=Y_{1}(\lambda)+Y^{\prime}(\lambda),
 \label{25}
%
\end{eqnarray}
where
%
\begin{eqnarray}
Y_{1}(\lambda)&=&-\frac{1}{\beta}\sum\limits_{\overrightarrow{k},\sigma,
\omega_{n}}[\ln(\epsilon(\overrightarrow{k})
\lambda\Lambda_{\lambda}(k)-1)\nonumber\\
&+&T_{\lambda}(k)\lambda\Lambda_{\lambda}(k)]\exp(i\omega_{n}0^{+}),
\label{26}
%
\end{eqnarray}
and $Y^{\prime}(\lambda)$ is constructed from skeleton diagrams
without external lines and is depicted on Fig.4
%
\begin{figure*}[t]
%
\centering
\includegraphics[width=0.85\textwidth,clip]{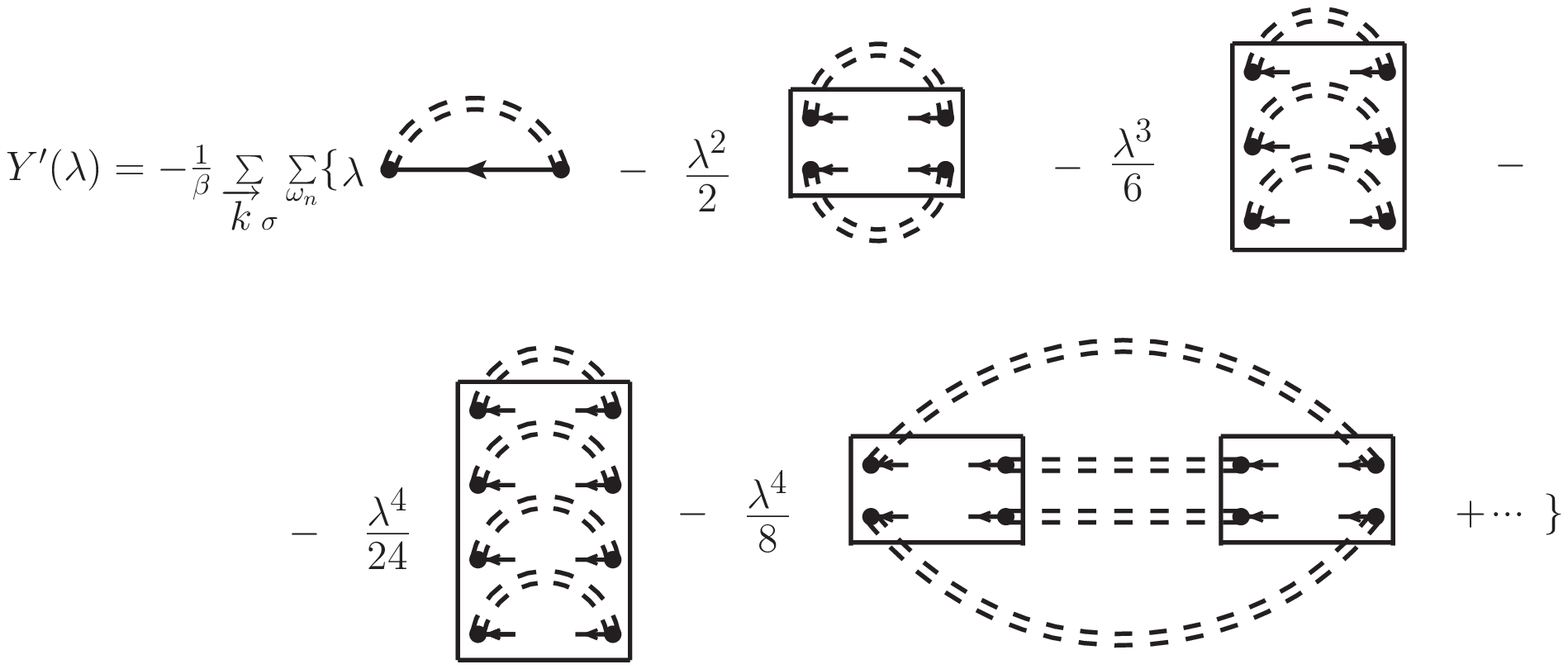}
\vspace{-0mm}
%
\caption{ The simplest skeleton diagrams for functional
$Y^{\prime}(\lambda)$. Double dashed lines are tunneling functions
$T_{\lambda}(k)$.} \label{fig-4} \vspace{-5mm}
\end{figure*}
%

The dependence on $\lambda$ in the functional
$Y^{\prime}(\lambda)$ is twofold, just through the dependence
of the renormalized Green's functions
$G_{\lambda},T_{\lambda},\Lambda_{\lambda}$ and
$\Sigma_{\lambda}$ and through the explicit factors
$\lambda^{n}$ in front of each diagram of
$Y^{\prime}(\lambda)$.

By using these definitions we can prove the equations
%
\begin{eqnarray}
\frac{\delta\beta Y_{1}(\lambda)}{\delta T_{\lambda}(k)}&=&-
\lambda\Lambda_{\lambda}(k)=-\Sigma_{\lambda}(k),\nonumber\\
\frac{\delta\beta Y^{\prime}(\lambda)}{\delta T_{\lambda}(k)}&=&
\lambda\Lambda_{\lambda}(k)=\Sigma_{\lambda}(k), \label{27}
%
\end{eqnarray}
and as a result we obtain the stationary property:
%
\begin{eqnarray}
\frac{\delta\beta Y(\lambda)}{\delta T_{\lambda}(k)}=0. \label{28}
%
\end{eqnarray}
By using the definition (13) of the mass operator
$\Sigma_{\lambda}(k)$ we can rewrite the functional
$Y_{1}(\lambda)$ in the form
%
\begin{eqnarray}
Y_{1}(\lambda)&=&-\frac{1}{\beta}\sum\limits_{\overrightarrow{k}\sigma}\sum\limits_{
\omega_{n}}[\ln(\epsilon(\overrightarrow{k})
\Sigma_{\lambda}(k)-1)\nonumber\\
&+&T_{\lambda}(k)\Sigma_{\lambda}(k)]\exp(i\omega_{n}0^{+}),
\label{29}
%
\end{eqnarray}
and prove the second form of stationary property
%
\begin{eqnarray}
\frac{\delta\ Y(\lambda)}{\delta \Sigma_{\lambda}(k)}=0,
\label{30}
%
\end{eqnarray}
if we take into account the Dyson equation for tunneling
function $T_{\lambda}(k)$.

Now we shall discuss the derivative by $\lambda$ of the functional
$Y(\lambda)$. As was mentioned above there is twofold dependence
of $\lambda$ through $\Sigma_{\lambda}(k)$ and explicit through
the factors $\lambda^{n}$ before skeleton diagrams of
$Y^{\prime}(\lambda)$ Fig. 4.

Due to the stationary property (30) we obtain
%
\begin{eqnarray}
\frac{dY(\lambda)}{d\lambda}&=&\sum\limits_{k}\frac{\delta
Y(\lambda)}{\delta\Sigma_{\lambda}(k)}\frac{\delta\Sigma_{\lambda}(k)}{\delta\lambda}+
\frac{\partial Y(\lambda)}{\partial\lambda}|_{\Sigma_{\lambda}}\\
&=&\frac{\partial
Y^{\prime}(\lambda)}{\partial\lambda}|_{\Sigma_{\lambda}}.\nonumber
\label{31}
%
\end{eqnarray}
Here we have taken into account the equation (30) and that that
$Y_{1}(\lambda)$ has not explicit dependence of $\lambda$.

By using Fig. 4 for $Y^{\prime}(\lambda)$ and the definition of
$\Lambda_{\lambda}$ of Fig. 2 it is easy to establish the
property:
%
\begin{eqnarray}
\lambda\frac{\partial
Y^{\prime}(\lambda)}{\partial\lambda}|_{\Sigma_{\lambda}}&=&\frac{1}{\beta}\sum\limits_{\overrightarrow{k}\sigma}\sum\limits_{
\omega_{n}}T_{\lambda}(k)\lambda\Lambda_{\lambda}(k)\exp(i\omega_{n}0^{+})\nonumber\\
&=&\frac{1}{\beta}Tr(T_{\lambda}\Lambda_{\lambda}). \label{32}
%
\end{eqnarray}
Therefore from (31) and (32) we obtain:
%
\begin{eqnarray}
\lambda\frac{d
Y(\lambda)}{d\lambda}&=&\frac{1}{\beta}\sum\limits_{\overrightarrow{k}\sigma}\sum\limits_{
\omega_{n}}T_{\lambda}(k)\Sigma_{\lambda}(k)\nonumber\\
&=&\frac{1}{\beta}Tr(T_{\lambda}\Sigma_{\lambda}). \label{33}
%
\end{eqnarray}
From equation (23) and (33) we have
%
\begin{eqnarray}
\lambda\frac{d F(\lambda)}{d\lambda}=\lambda\frac{d
Y(\lambda)}{d\lambda}, \label{34}
%
\end{eqnarray}
and as the consequence we obtain
%
\begin{eqnarray}
F(\lambda)= Y(\lambda)+\textbf{const.} \label{35}
%
\end{eqnarray}
Because for $\lambda=0$ perturbation is absent and
$F(0)=F_{0}$,$Y(0)=0$ we have
%
\begin{eqnarray}
F(\lambda)= Y(\lambda)+F_{0}. \label{36}
%
\end{eqnarray}
Now we can put $\lambda=1$ and to obtain
%
\begin{eqnarray}
F(1)= Y(1)+F_{0}. \label{37}
%
\end{eqnarray}
with the stationary property
%
\begin{eqnarray}
\frac{\delta F}{\delta \Sigma}=0. \label{38}
%
\end{eqnarray}

\section{Conclusions}


We have developed the diagrammatic theory proposed for Hubbard model
many years ago and introduced the notion of renormalized tunneling
Green's function $T$ in addition to known before. We have defined
correlation function $\Lambda$, the mass operator $\Sigma$ for
tunneling function and establish the Dyson equation for them. The
mass operator was found (13) to be equal to correlation function for
$\lambda=1$.

We have obtained the diagrammatic representation of the
correlation function through the skeleton diagrams which contain
the many-particle irreducible Green's functions $G_{2n}^{(0)ir}$
with all possible values of $n-$ order of perturbation theory and
the renormalized tunneling Green's functions.

We have established the relation between renormalized
quantities of thermodynamic potential and one-particle Green's
function. This last function has additional dependence on
auxiliary constant $\lambda$ and has to be integrated over
it.

We have proved the possibility to avoid such integration by
$\lambda$ and to introduce the special functional $Y(\lambda)$
constructed from skeleton diagrams.

At the final part of the paper we have proved the stationary
property of this functional and found its relation to stationary of
thermodynamic potential with respect variation of mass operator or
full tunneling function. This theorem is the generalization of known
Luttinger and Word theorem $^{[15]}$ proved for weakly correlated
systems on the case of strongly correlated systems described with
the Hubbard model.
%
\begin{acknowledgments}
Two of us (V.A.M. and L.A.D.) would like to thank Professor N.M.
Plakida and Dr. S. Cojocaru for a very helpful discussion.
\end{acknowledgments}
%

%

\end{document}